\documentclass[10pt,reprint]{socc14}

\usepackage{amsmath}
\usepackage{mathptmx}

\usepackage{graphicx,mathrsfs,amsfonts}

\usepackage{balance}
\usepackage[dvipdfm,CJKbookmarks=true,bookmarksopen=true,colorlinks,citecolor=black,linkcolor=black,anchorcolor=black,urlcolor=black] {hyperref}

\begin{document}

\setlength{\pdfpageheight}{\paperheight}
\setlength{\pdfpagewidth}{\paperwidth}

%\conferenceinfo{Submission to SoCC '15}{August, 2015, Hawai'i, USA}
\copyrightyear{2015}
\copyrightdata{978-1-nnnn-nnnn-n/yy/mm}
\doi{nnnnnnn.nnnnnnn}

% Uncomment one of the following two, if you are not going for the
% traditional copyright transfer agreement.

%\exclusivelicense                % ACM gets exclusive license to publish,
                                  % you retain copyright

%\permissiontopublish             % ACM gets nonexclusive license to publish
                                  % (paid open-access papers,
                                  % short abstracts)

%\titlebanner{banner above paper title}        % These are ignored unless
%\preprintfooter{short description of paper}   % 'preprint' option specified.

\title{\Large \bf 10 Observations on Google Cluster Trace +\\ 2 Measures for Cluster Utilization Enhancement}
%big data about service/latency
%the bigness in the tail

\authorinfo{Yuqing Zhu, Yilei Wang, Fan Wang}%, Yilei Wang, Fan Wang
           {Institute of Computing Technology, Chinese Academy of Sciences}
           {zhuyuqing@ict.ac.cn}%,wangyilei,wangfan

\maketitle

\begin{abstract}
Utilization enhancement is a key concern to cluster owners. Google's cluster manager named Borg manages its clusters at an overall utilization higher than many others' clusters. Recently, Google has disclosed the details of its powerful cluster manager Borg. Quite a few lessons are summarized from the Borg experiences. Nevertheless, we find that more can be learned if the Borg design is correlated with the trace analysis of a Google cluster managed by Borg. There is one such trace released four years ago. In this paper, we analyze the Google cluster trace and make 10 observations not found in previous analyses. We also correlates the results of our analysis and previous analyses to the Borg design, such that we find two measures that can possibly further improve cluster utilization over Borg.
\end{abstract}

\section{Introduction}

As Internet and Big Data are playing a more and more important role, many entities deploy clusters of computers for supporting their own services, as well as storing and analyzing their private data. Fearing to impair the SLA (service level agreement) of their services, cluster owners usually use separate clusters for online service and data analysis. This leads to low overall utilization of clusters. A slight enhancement of cluster utilization can mean much fewer machines and much lower costs. Therefore, cluster owners seek to improve cluster utilization by every means.

Recently, Google publishes a paper on its power cluster manager Borg~\cite{borg}, which effectively manages Google clusters at an overall utilization higher than many others' clusters. Borg has exploited a bunch of techniques, including the notable ones of cluster sharing~\cite{clustersharing}, task packing~\cite{packing}, resource reclamation, and fine-grained resource scheduling.

The revealment of the above techniques advance our understanding of a previously disclosed Google cluster trace~\cite{googledata}. For example, short-lived tasks account for less than 10\% of utilization~\cite{reisssocc12}. While one expecting short-lived tasks are latency sensitive services, these tasks are actually components of no-production jobs. In fact, latency-sensitive production services are mostly deployed as long-running tasks, taking up a large portion of utilization. It is cluster sharing that enables the consolidation of latency-sensitive production services and non-production batch jobs in a common shared cluster.

Another example of new understandings is about why the sum of all allocated CPU at any moment exceeds the total CPU capacity of the cluster. Generally, it is dangerous to run latency-sensitive services in such a way because these services can have utilization spikes even overconsuming the allocated resources. With cluster sharing and task packing, latency-sensitive service tasks run on the same machine with non-production tasks, which can be safely evicted. When over-utilization happens, the mechanism of resource reclamation enables short-term reclaiming of allocated resources used by short-lived low-priority tasks.

In this paper, we propose to further the understanding of cluster utilization enhancement by correlating the Google cluster trace~\cite{googledata} to the design of the Google cluster manager, i.e. Borg~\cite{borg}. We process the trace data in a way different from previous analyses~\cite{reisssocc12, googleenergy12, googleliu12, googlegrid12}. We carry out heavy computation on the trace data. The trace includes data on jobs, task and machines. We focus on tasks in our analysis because task is the basic unit for scheduling and running. Besides, properties of tasks directly relate to resource utilization. We join the large tables of task and jobs. We also aggregate on several key global properties of the cluster for all time moments.

The differed way of our processing enables us to find ten new observations not found before, concerning task submission, completion, and scheduling patterns, as well as task execution times. We also discover new task transitions not described in the specification of the trace. We correlate results of our analysis and previous analyses to the Borg design such that we find two measures that can possibly further improve cluster utilization over Borg.

Next, we first overview the four scheduling techniques of Borg and the released Google cluster trace in Section \ref{sec:background}. Then, we illustrate our 10 observations one by one in Section \ref{sec:observe}. We analyze and summarize the two measures for cluster utilization enhancement in Section \ref{sec:measures}. Finally, we conclude the paper in Section \ref{sec:conclude}.

\section{Background}
\label{sec:background}

\subsection{Four Scheduling Techniques of Borg}

Borg has exploited a bunch of techniques. Among these techniques, four are most notable and closely related to the analysis in this paper. They are cluster sharing~\cite{clustersharing}, task packing~\cite{packing}, resource reclamation, and fine-grained resource scheduling~\cite{borg}

\textbf{Cluster sharing} has been considered by many as can drastically increase the cost of running programs. However, the Borg paper shows that this does not necessarily happen, if proper security and performance isolation are provided. Even if cluster sharing can lead to CPU slowdown, the effect is outweighed by the decrease of machine number. \textbf{Task packing} requires the scheduler to pack tasks into as fewer machines as possible. This will result in more flexibility for later scheduling and higher utilization. Using \textbf{resource reclamation}, a cluster manager can reclaim the allocated computing resources from running tasks, which usually use only a portion of the requested or the allocated resources. Resource reclamation is a very effective technique to improve cluster utilization. It relies on resource estimation and real-time monitoring. \textbf{Fine-grained resource scheduling} enables the effective implementation of the above three techniques.

\subsection{Google Cluster Trace}

Google has released a one-month trace for one of its consolidated clusters~\cite{googledata}. The trace includes six components, covering machine attributes, machine events, job events, task constraints, task events, and task properties. Notably, task properties include information of real-time resource utilization.%28.9 days

The Google cluster represented by the trace is constructed from a variety of machine classes with varied processing capacities. The workload in the trace demonstrates a high degree of heterogeneity and variability. It is composed of jobs with varied resource requests and processing priorities. Each job is split into one or more tasks, which is the basic unit for resource allocation and scheduling. Tasks inherit the processing priority and the resource request from the job. A few early analyses on the trace have revealed some facts, e.g. periodical pattern and high dynamism, but not including our findings.

Difficulty exists in processing the big data of the trace. The data volume of the trace is about 40 gigabytes \emph{after compression}. The size of all tables after decompression is about 300GB. To obtain our findings, multiple joins must be carried out on the six components of the trace. We tried the common approach by using MySQL to analyze the data, but a join does not produce any result even after \emph{a week}. Thus, we write our own programs to do the analysis.
\begin{figure}[!t]
%\renewcommand{\captionfont}{\Large \bfseries \sffamily} \renewcommand{\captionlabelfont}{}
    %\vspace{-12pt}
      \centering
      \includegraphics[width=0.48\textwidth]{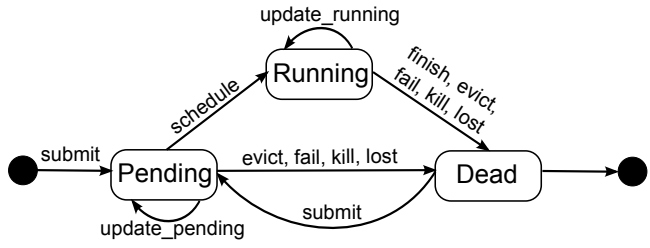}
      \caption{State transitions for jobs and tasks.}%\vspace{-12pt}
      \label{fig:transitions} %% label for entire figure
\end{figure}

\textbf{State Transition Graph.} A task submitted to the system can be in one of the three states, i.e. pending, running and dead. The three states are represented by the rounded rectangles depicted in Figure \ref{fig:transitions}. Actions that trigger the state transitions are listed on the arrows between states. Our analysis show that the actual transition graph for the trace differs from the one depicted in the Borg paper~\cite{borg} and that in the data description document~\cite{googledataformat}. Two differences exist. First, a submitted task can be evicted directly from the pending state. Second, a submitted task cannot be resubmitted directly, i.e. that a \emph{submit} event cannot happen right after another \emph{submit} event for a task. Figure \ref{fig:transitions} is the state transition graph updated with our analysis results. In the following, our analysis is based on this state transition graph.

\section{Observations}
\label{sec:observe}

In this section, we illustrate ten new observations based on the Google trace. We divide the observations into three groups. The first group of observations shows that Borg is very effective in controlling the processing workloads of the cluster. The second group shows that tasks are cumulating in the cluster for the whole month. The third group shows that the time for task execution and scheduling is not affected by the increasing number of tasks in the cluster.
\begin{figure}[!t]
%\renewcommand{\captionfont}{\Large \bfseries \sffamily} \renewcommand{\captionlabelfont}{}
    %\vspace{-12pt}
      \centering
      \includegraphics[width=0.48\textwidth]{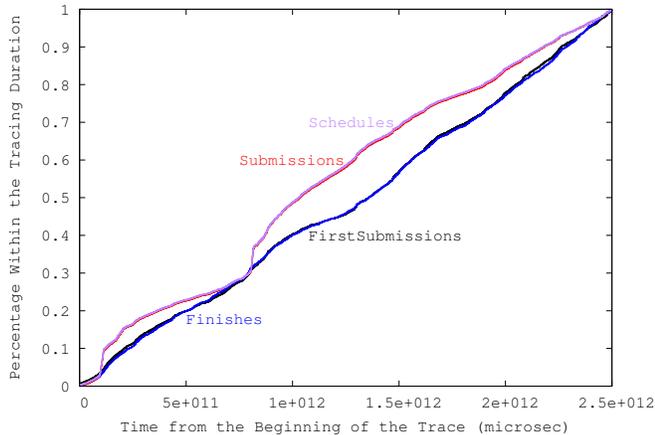}
      \caption{The cumulative distributions of new task submissions, task completions, task submissions and task scheduling within the tracing period of a month.}\vspace{-6pt}
      \label{fig:0a} %% label for entire figure
\end{figure}

\subsection{Effective Cluster Manager and Scheduler}

\paragraph{Analysis method.}

For the whole tracing period of a month, we plot the cumulative distribution function (CDF) graph regarding the number of new task submissions for every microsecond in Figure \ref{fig:0a}. To make the plot, we keep the list of unique tasks and traverse all records of task events in the order of time. We only count the number of tasks that are submitted for the first time for each timestamp, excluding those that are resubmitted.

Similarly, we plot the CDF graph regarding the number of task completions, task submissions and scheduled tasks for every microsecond in Figure \ref{fig:0a}. Note that \emph{task submissions} differs from \emph{new task submission} in that the former also includes task resubmissions.

\paragraph{Observation 1: The rate of new task submissions is stable.}

Strikingly, the CDF graph of new task submissions can be approximated by a line. The shape of this CDF graph contradicts from our understanding that event arrivals normally follow the Poisson distribution. We think the contradiction is unlikely, as the trace includes a month's records for every microsecond. Thus, we conjecture that the cluster manager can keep the number of new task submissions in a static rate. As the cluster is consolidating production jobs with non-production jobs, the cluster manager can effectively control the submission of non-production jobs.

\paragraph{Observation 2: The rate of task completions is also stable.}

Notably, the CDF graph of task completions can also be approximated by a line. Besides, it also coincides with the CDF graph of new task submissions. We are unaware whether this patten is common for consolidated cluster environments, or it is because of Google's typical cluster manager. But from the coincidence of two graphs, we believe that the effective control of task submissions by Borg results in the CDF graph of the task completions.

\paragraph{Observation 3: Task resubmissions are concentrated at two time periods.}

The CDF graph of task submissions superposes the CDF graph of new task submissions at some time periods, while the former exhibits two lumps with regard to the latter. Considering that task submissions include the resubmissions of tasks, we owe the two lumps to two peaks of task resubmissions. This phenomenon is similar to the phenomenon of periodical task submissions revealed in an early analysis~\cite{reisssocc12}.
\begin{figure}[!t]
%\renewcommand{\captionfont}{\Large \bfseries \sffamily} \renewcommand{\captionlabelfont}{}
    %\vspace{-12pt}
      \centering
      \includegraphics[width=0.46\textwidth]{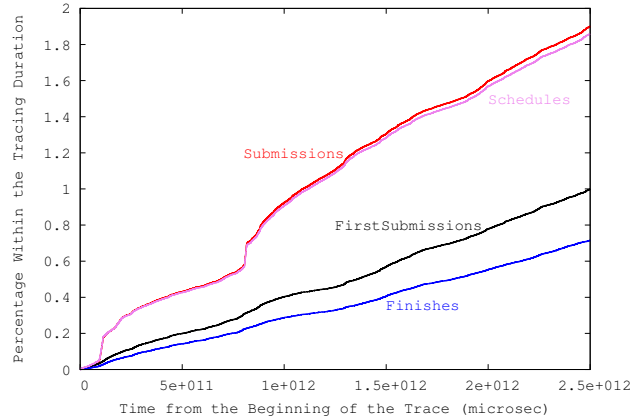}
      \caption{Weighted cumulative distributions of new task submissions, task completions, task submissions and task scheduling within the tracing period of a month.}\vspace{-6pt}
      \label{fig:0b} %% label for entire figure
\end{figure}

\paragraph{Observation 4: Task scheduling tightly follows task submissions.}

The CDF graph of scheduled tasks superposes that of task submissions. The natural reasoning on this superposition is that the scheduler of the Google cluster can effectively schedule all submitted tasks.

\subsection{Task Cumulation}

To further investigate the superposition phenomenon of CDF graphs, we answer two related questions in this section: (1) how much of the newly submitted tasks are completed; (2) whether all submitted tasks get scheduled in time. The CDF graphs in the previous section only disclose the general trend of the corresponding events. They cannot answer these questions.

\paragraph{Analysis method 3-1.}

We plot weighted CDF graphs. We let the CDF graph of new task submission be the standard graph. Then, we multiply the data points of the other three graphs by a weight. The weight is the ratio of the total number of the corresponding events to that of new task submissions. For example, all data points of the task completion CDF graph are multiplied by the ratio of the total number of the task completions to the total number of new task submissions.

Thus, we count the total numbers of new task submissions, task completions, task submissions and task scheduling events. The total number of new task submissions, task completions, task submissions and task scheduling events are 2.5424731E7, 1.8217975E7, 4.8375166E7 and 4.7331507E7 respectively. We compute the weights based on these numbers and draw the weighted CDF graphs in Figure \ref{fig:0b}.

\begin{figure}[!t]
%\renewcommand{\captionfont}{\Large \bfseries \sffamily} \renewcommand{\captionlabelfont}{}
    %\vspace{-12pt}
      \centering
       \includegraphics[width=0.48\textwidth]{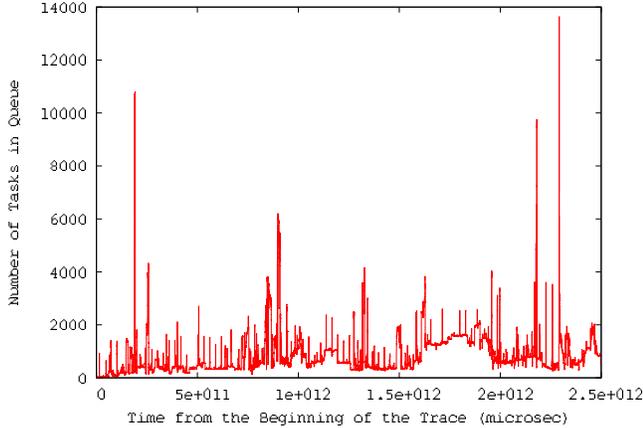}
      \caption{The number of queuing tasks as a function of time during the tracing period of a month.}\vspace{-6pt}
      \label{fig:0c} %% label for entire figure
\end{figure}

\paragraph{Observation 5: Unfinished tasks are cumulating in the cluster.}

The rate of task completions is smaller than that of new task submissions. Thus, the number of unfinished tasks are increasing in the cluster. These unfinished tasks can be lost, failed, killed and evicted to leave the system. We find that this is not the truth, though. From the graph of task submissions, which include resubmitted tasks, we find that most of the unfinished tasks are resubmitted to the cluster. Some are even resubmitted again and again, leading to the increasing number of total submissions. As the rate of task completions are stable, the unfinished tasks are actually cumulating in the cluster.

\paragraph{Observation 6: Tasks can generally be scheduled in time, but the schedulers cannot catchup at the end of the tracing period.}

The weighted CDF graphs of task submissions and scheduling events still superimpose each other. However, they diverge at the end of the tracing period. The number of task submissions are increasingly exceeding the number of scheduled tasks. That is, the schedulers are not scheduling tasks as fast as tasks are submitted and resubmitted to the cluster.

To verify this observation, we analyze the number of tasks queueing to be scheduled, as well as the number of tasks being processed.

\paragraph{Analysis method 3-2.}

We plot the number of tasks queueing to be scheduled in the cluster as a function of the time. To make the plot, we group all task events by job ID and task index, the two of which are the unique identifier of tasks. Within each task group, we sort events in time order. Then, we sort out the graph of state transitions for each task. The possible task transitions are depicted in Figure \ref{fig:transitions}.

We further process the transition graph to collapse all edges involving \emph{update\_pending} or \emph{update\_running}. The processed transition graph contains no \emph{update\_pending} or \emph{update\_running} events. After processing, a \emph{submit} event can only head to \emph{schedule}, \emph{fail}, \emph{kill} or \emph{lost} events. Similarly, a \emph{schedule} event can only head to \emph{finish}, \emph{evict}, \emph{fail}, \emph{kill} or \emph{lost} events.

We further sort the processed transition graph according to the time order. Then, we aggregate the following numbers for each timestamp: submitted and scheduled events, as well as \emph{fail(pFail)}, \emph{kill(pKill)} and \emph{lost(pLost)} events with \emph{submit} event as predecessor. For each timestamp, we compute the change of numbers of tasks queueing to be scheduled (denoted as \emph{queueNumChange}) as the number of submitted tasks minus the number of scheduled, \emph{pFail}, \emph{pKill}, and \emph{pLost} tasks.

The number of tasks queueing to be scheduled for a timestamp is the summation of \emph{queueNumChange} for all past timestamps and the current timestamp. The initial number of tasks queueing to be scheduled is zero.
\begin{figure}[!t]
%\renewcommand{\captionfont}{\Large \bfseries \sffamily} \renewcommand{\captionlabelfont}{}
    %\vspace{-12pt}
      \centering
        \includegraphics[width=0.48\textwidth]{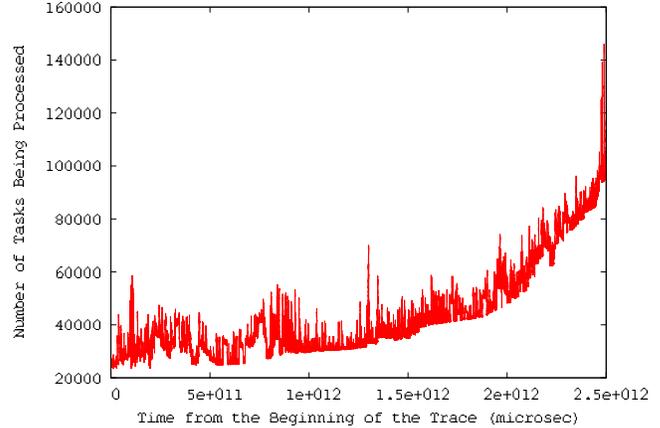}
      \caption{The number of running tasks as a function of time during the tracing period of a month.}\vspace{-6pt}
      \label{fig:0d} %% label for entire figure
\end{figure}

\paragraph{Observation 7: The number of tasks queueing to be scheduled is slightly increasing.}

Figure \ref{fig:0c} is the graph for the number of tasks queueing to be scheduled as a function of microseconds. It is obvious that the number is increasing as time passes. There are more and more tasks waiting to be scheduled as the cluster works for longer time. That is, the scheduler of the system is either overloaded, or it is restricting the number of scheduled tasks on purpose.

We continue to analyze how the number of in-execution tasks is changing in the cluster.

\paragraph{Analysis method 3-3.}

We continue to analyze the number of in-execution tasks. The initial number of running tasks is zero. The number of in-execution tasks, i.e. running tasks, is the summation of \emph{runningNumChange} for all past timestamps and the current timestamp. For each timestamp, we compute the change of numbers of running tasks (denoted as \emph{runningNumChange}) as the number of scheduled tasks minus the number of \emph{finished}, \emph{evict}, \emph{rFail}, \emph{rKill} and \emph{rLost} tasks. Based on the transition graph in Figure \ref{fig:transitions}, we aggregate the following numbers for each timestamp: \emph{finish} and \emph{evict} events, as well as \emph{fail(rFail)}, \emph{kill(rKill)} and\emph{ lost(rLost)} events with schedule event as predecessor.

\paragraph{Observation 8: The number of in-execution tasks is surging.}

Figure \ref{fig:0d} is the graph of the number of in-execution tasks as a function of microseconds. The number is also increasing as time passes. As the time heads towards the end of the tracing period, the number surges. The reason can be found from Figure \ref{fig:0b}: the number of scheduled tasks far exceeds the number of completed tasks.

\subsection{Unaffected Task Execution and Scheduling Time}
\begin{figure}[!t]
%\renewcommand{\captionfont}{\Large \bfseries \sffamily} \renewcommand{\captionlabelfont}{}
    %\vspace{-12pt}
      \centering
      \includegraphics[width=0.48\textwidth]{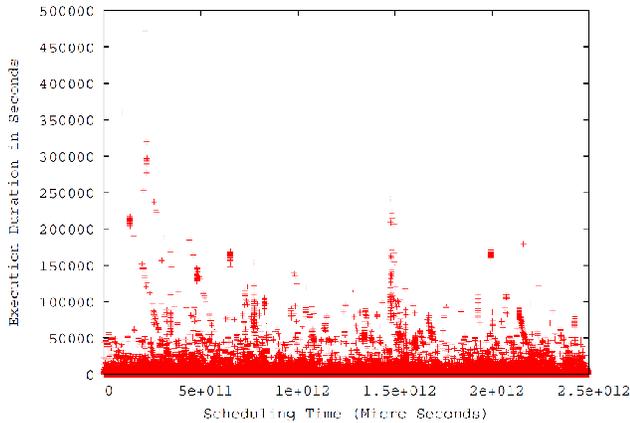}
      \caption{Moving average of task execution time during the tracing period of a month.}\vspace{-6pt}
      \label{fig:0e} %% label for entire figure
\end{figure}
\begin{figure}[!b]
   \vspace{-6pt}
      \centering
      \includegraphics[width=0.48\textwidth]{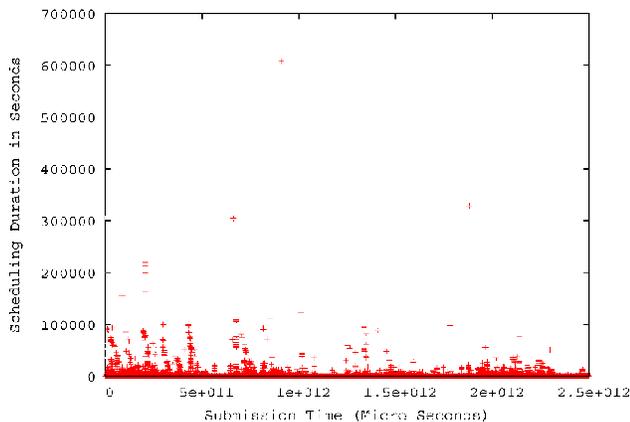}
      \caption{Moving average of task scheduling time during the tracing period of a month.}%\vspace{-12pt}
      \label{fig:0f} %% label for entire figure
\end{figure}

In this section, we investigate whether the increasing numbers of queueing and under-processing tasks can affect the processing time or the scheduling time of tasks. Thus, we plot the moving average of task execution and scheduling times for each timestamp respectively. The resulting graphs are Figure \ref{fig:0e} and Figure \ref{fig:0f}.

\paragraph{Observation 9: Task execution and scheduling times are not influenced by the cumulation of tasks.}

Figure \ref{fig:0e} and Figure \ref{fig:0f} demonstrate huge variances. Some tasks have a long execution time. Such tasks can be the latency-sensitive services, which initiate long running tasks with high priority to guarantee user-level satisfaction, as described in the Borg paper. Some tasks have a long scheduling time. The underlying reason is very likely to be that no computing units with matching resources can satisfy the task requests. This makes the pending tasks keep updating its constraints.

\paragraph{Observation 10: Latency-sensitive tasks are usually long-running tasks.}

\begin{figure}[!t]
%\renewcommand{\captionfont}{\Large \bfseries \sffamily} \renewcommand{\captionlabelfont}{}
    %\vspace{-12pt}
      \centering
      \includegraphics[width=0.45\textwidth]{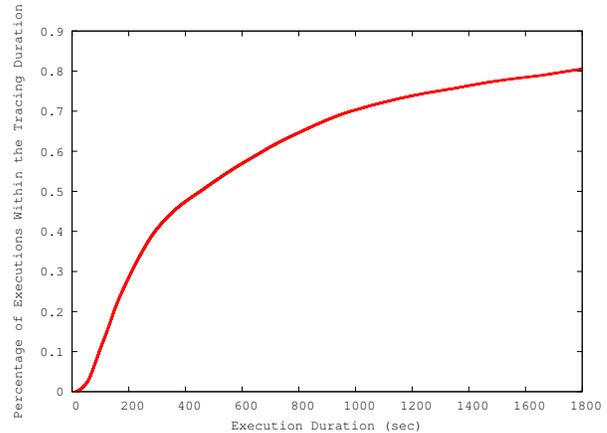}
      \caption{CDF of task execution durations.}\vspace{-12pt}
      \label{fig:taskExecTimeDist} %% label for entire figure
\end{figure}

Tasks execution times last from 1 microsecond to 8.969 days. From the paper of Borg, we know that latency-sensitive services can only tolerate processing times shorter than a few hundred milliseconds. Such services are long running tasks. We find that only 18,217,749 of the 25,444,397 tasks are actually executed to completion. Among the finished tasks, only 2 tasks have execution times shorter than 1 second. The tasks with the third shortest execution time run for more than 8 seconds. About 80\% of tasks have an execution time shorter than 30 minutes, as shown in Figure \ref{fig:taskExecTimeDist}.
%The total number of tasks is 25,444,397.

%
\section{Two Measures to Improve Utilization}
\label{sec:measures}

In this section, we first summarize the utilization facts found in previous analyses. By correlating our findings and the utilization facts to the Borg design, we present two measures for cluster utilization enhancement.

\subsection{Preliminary: Utilization Facts}
\begin{figure}[!b]
    \vspace{-12pt}
      \centering
      \includegraphics[width=0.46\textwidth]{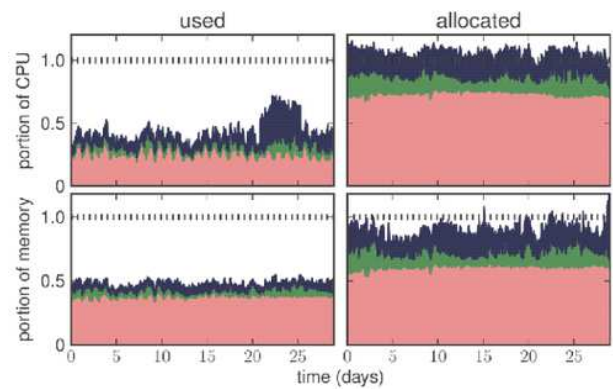}
      \caption{Moving hourly average of CPU (top) and memory (bottom) utilization (left) and resource requests (right). Stacked plot by priority range, highest priorities (production) on bottom (in red/lightest color), followed by the middle priorities (green), and gratis (blue/darkest color). The dashed line near the top of each plot shows the total capacity of the cluster (cited from~\cite{reisssocc12}).}%\vspace{-12pt}
      \label{fig:utilization} %% label for entire figure
\end{figure}
\begin{figure}[!t]
%\renewcommand{\captionfont}{\Large \bfseries \sffamily} \renewcommand{\captionlabelfont}{}
    %\vspace{-12pt}
      \centering
      \includegraphics[width=0.468\textwidth]{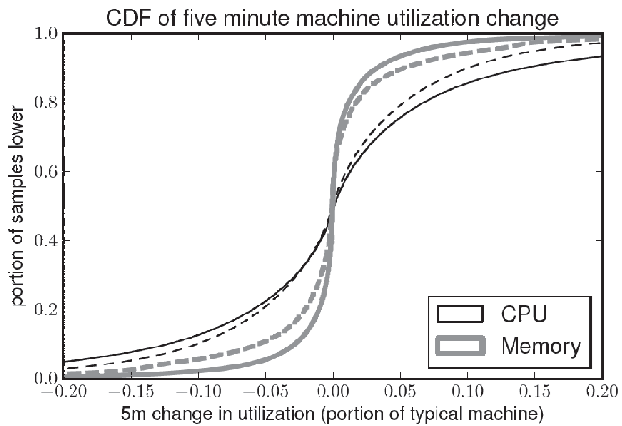}
      \caption{CDF of changes in average machine utilization between two consecutive five minute sampling periods. Solid lines exclude tasks which start or stop during one of the five minute sampling periods (cited from~\cite{reisssocc12}).}\vspace{-6pt}
      \label{fig:utilizationChange} %% label for entire figure
\end{figure}
Tasks are tagged with priorities. Production tasks are those with priority values greater or equal to 9. The total number production tasks is 1,558,255, occupying 6\% of all tasks. Production tasks are latency sensitive and should not be ¡°evicted due to over-allocation of machine resources¡±~\cite{googledataformat}. This small portion of tasks uses more than 80\% of CPU and memory, as can be seen from Figure \ref{fig:utilization}.

The initial impression of Figure \ref{fig:utilization} is that almost all tasks are claiming much more resources than they actually need. However, Figure \ref{fig:utilization} only represents an overall phenomenon. For individual tasks, they have spikes of resource consumptions that can well exceed their allocated quota~\cite{reisssocc12,borg}. The overconsumption of resources can be as much as twice of the requested ones. In other words, most users are not necessarily always over-requesting resources, but the overall cluster utilization does not look optimized.

Now consider the utilization changes of the cluster in Figure \ref{fig:utilizationChange}. The Google cluster sets a utilization sampling period of 300 seconds~\cite{googledataformat,borg}, i.e. 5 minutes. Between two consecutive sampling periods, the consumed CPU or memory do not change abruptly. For more than 90\% of sampling moments, the consumption fluctuates in a range smaller than 20\%. With all these utilization facts, we can deduce the following utilization improvement measures by closely considering the Borg design.

\subsection{Two Measures}

The following measures must be considered basing on the Borg design, especially the four techniques of cluster sharing, task packing, resource reclamation, and fine-grained resource scheduling.

\paragraph{1. The limited fluctuation of utilization in consecutive moments makes possible setting a narrow margin for resource reservation.} Hence, more tasks can be assigned to the same resources, although more non-production tasks might be evicted due to increased ad-hoc resource demands by production tasks.\\

The resource reclamation in Borg depends on resource reservation, which reserves resources equal to the actual usage plus a safety margin. This is done through estimation, which is done through shrinking and decaying from the requested size of resources. Thus, excessive resources are wasted before the estimation shrinks to proper values.

As shown in Figure \ref{fig:utilizationChange}, the consumption of CPU or memory do not change abruptly in the cluster. More than 90\% of times have changes less than 20\% between consecutive sampling periods. As a result, there is high confidence in that setting the margin of resource reservation to 20\% of a previous utilization is enough. This margin can be raised according to the utilization change distributions, if tighter SLA is required.

Now, consider the CPU utilization and allocation rates in Figure \ref{fig:utilization}. The utilization of production tasks only takes up about 50\% of the allocated resources. That is, the average margin is about 100\%. Using utilization change distributions to set the margin should improve the utilization results.

On the other hand, we have noted that the number of tasks in queue and being processed is increasing in the cluster (Figure \ref{fig:0c} and \ref{fig:0d}). The Borg report points out that CPI is a key factor for performance, and CPI is closely related to the number of running tasks and the CPU rate on a machine~\cite{borg}. We believe that the task cumulation phenomenon must be considered along with resource reservation, so as to avoid drastic performance degradation.

\paragraph{2. Sampling utilizations at finer time periods to enable setting even narrower margins for resource reservation.} Hence, even more tasks can be assigned to the same resources.\\

Although a sampling period of 5 minutes produce a good result, a shorter sampling period can model the utilization change behavior more faithfully, such that the resource reservation can be more precise. On the other hand, only two tasks run for shorter than 8 seconds (Figure \ref{fig:taskExecTimeDist}), indicating that the sampling period need not to be finer than 8 seconds.

\begin{figure}[!t]
%\renewcommand{\captionfont}{\Large \bfseries \sffamily} \renewcommand{\captionlabelfont}{}
    %\vspace{-12pt}
      \centering
      \includegraphics[width=0.48\textwidth]{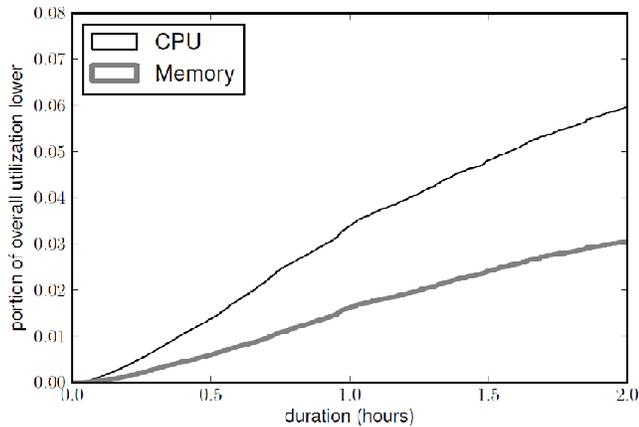}
      \caption{CDF of utilization by task duration. Note that tasks running for less than two hours account for less than 10\% of utilization by any measure (cited from~\cite{reisssocc12}).}%\vspace{-12pt}
      \label{fig:utilizationShortTask} %% label for entire figure
\end{figure}

Furthermore, short-lived tasks actually use a tiny portion of total CPU resources (Figure \ref{fig:utilizationShortTask}). And, about 40\% of tasks have an execution time shorter than 5 minutes (Figure \ref{fig:taskExecTimeDist}). If short-lived tasks always over-request resources, the overall utilization of a cluster can drop fiercely, regardless of using the decaying resource reservation or the change-based resource estimation. A finer sampling periods is useful to prevent such cases. What is even better is to sample and model utilization changes for different types of tasks, e.g. long-running and short-lived tasks.

\section{Conclusion}%
\label{sec:conclude}

In this paper, we analyze a Google cluster trace and correlate the analysis results with the design of the Google cluster manager Borg. We obtain ten new observations not found in previous analyses on the trace. We also deduce two measures that can possibly further improve cluster utilization over Borg. First, the limited fluctuation of utilization in consecutive moments makes possible setting a narrow margin for resource reservation. Second, sampling utilizations at finer time periods to enable setting even narrower margins for resource reservation.

In the future, we will continue this line of work and evaluate the efficiency of the two measures. Our work also points to a direct research problem about how to quickly obtain the utilization change distributions of a large cluster to make the effective real-time resource estimation possible.

\section{Acknowledgments}

This work is supported in part by the State Key Development Program for Basic Research of China (Grant No. 2014CB340402) and the National Natural Science Foundation of China (Grant No. 61303054).

\balance

{\footnotesize
\bibliographystyle{abbrvnat}
\bibliography{ref}

\begin{thebibliography}{9}
\providecommand{\natexlab}[1]{#1}
\providecommand{\url}[1]{\texttt{#1}}
\expandafter\ifx\csname urlstyle\endcsname\relax
  \providecommand{\doi}[1]{doi: #1}\else
  \providecommand{\doi}{doi: \begingroup \urlstyle{rm}\Url}\fi

\bibitem[Di et~al.(2012)Di, Kondo, and Cirne]{googlegrid12}
S.~Di, D.~Kondo, and W.~Cirne.
\newblock Characterization and comparison of cloud versus grid workloads.
\newblock In \emph{Cluster Computing (CLUSTER), 2012 IEEE International
  Conference on}, pages 230--238. IEEE, 2012.

\bibitem[Kambadur et~al.(2012)Kambadur, Moseley, Hank, and Kim]{clustersharing}
M.~Kambadur, T.~Moseley, R.~Hank, and M.~A. Kim.
\newblock Measuring interference between live datacenter applications.
\newblock In \emph{Proc. of Int¡¯l Conf. for High Performance Computing,
  Networking, Storage and Analysis (SC)}. IEEE, 2012.

\bibitem[Liu and Cho(2012)]{googleliu12}
Z.~Liu and S.~Cho.
\newblock Characterizing machines and workloads on a google cluster.
\newblock In \emph{Parallel Processing Workshops (ICPPW), 2012 41st
  International Conference on}, pages 397--403. IEEE, 2012.

\bibitem[Reiss et~al.(2011)Reiss, Wilkes, and Hellerstein]{googledataformat}
C.~Reiss, J.~Wilkes, and J.~L. Hellerstein.
\newblock Google cluster-usage traces: format+ schema.
\newblock \emph{Google Inc., White Paper}, 2011.

\bibitem[Reiss et~al.(2012)Reiss, Tumanov, Ganger, Katz, and
  Kozuch]{reisssocc12}
C.~Reiss, A.~Tumanov, G.~R. Ganger, R.~H. Katz, and M.~A. Kozuch.
\newblock Heterogeneity and dynamicity of clouds at scale: Google trace
  analysis.
\newblock In \emph{Proceedings of the Third ACM Symposium on Cloud Computing},
  page~7. ACM, 2012.

\bibitem[Verma et~al.(2014)Verma, Korupolu, and Wilkes]{packing}
A.~Verma, M.~Korupolu, and J.~Wilkes.
\newblock Evaluating job packing in warehouse-scale computing.
\newblock In \emph{Proc. of IEEE Cluster}. IEEE, 2014.

\bibitem[Verma et~al.(2015)Verma, Pedrosa, Korupolu, Oppenheimer, Tune, and
  Wilkes]{borg}
A.~Verma, L.~Pedrosa, M.~Korupolu, D.~Oppenheimer, E.~Tune, and J.~Wilkes.
\newblock Large-scale cluster management at google with borg.
\newblock In \emph{Proceedings of the Tenth European Conference on Computer
  Systems}, page~18. ACM, 2015.

\bibitem[Wilkes(2014)]{googledata}
J.~Wilkes.
\newblock More google cluster data. google research blog, November 2014.
\newblock https://code.google.com/p/googleclusterdata/wiki/ClusterData2011\_2.

\bibitem[Zhang et~al.(2012)Zhang, Zhani, Zhang, Zhu, Boutaba, and
  Hellerstein]{googleenergy12}
Q.~Zhang, M.~F. Zhani, S.~Zhang, Q.~Zhu, R.~Boutaba, and J.~L. Hellerstein.
\newblock Dynamic energy-aware capacity provisioning for cloud computing
  environments.
\newblock In \emph{Proceedings of the 9th international conference on Autonomic
  computing}, pages 145--154. ACM, 2012.

\end{thebibliography}
}

\end{document}